\documentclass[dvips]{article}
\usepackage[dvips]{graphicx}
\usepackage{epsfig}
\usepackage{amsfonts}
\usepackage{amsmath}
\usepackage{mathrsfs}
\usepackage{amssymb}
\usepackage{makeidx}
\usepackage{cancel}

\newtheorem{theorem}{Theorem}[section]
\newtheorem{definition}{Definition}[section]

\numberwithin{equation}{section}
\newcommand{\defeq}{\stackrel{\scriptscriptstyle{{\rm def}}}{=}}

\author{Jeffrey M. Groah}
\begin{document}
{\setlength\arraycolsep{1pt} 
\title{Riemannian Geometry of $C^{1,1}$ Manifolds}

\maketitle

\section{Introduction}

The Einstein equations, as a system of second-order partial differential equations, are expected to produce metric solutions that are $C^{1,1}$ and hence spacetime coordinates that are $C^{2,1}$. This level of regularity is sufficient for the existence of locally intertial frames and in \cite{Re-Te} such coordinates are shown to exist in the case of shockwave interactions between shocks from different characteristic families. See also \cite{Israel} for the case of a single shockwave and \cite{Op-Sn, Sm-TeNon}. In \cite{Gr-Te}, existence of shock wave solutions to the Einstein equations is given in the case of spherically symmetric spacetimes. In the case of spherically symmetric solutions of the Einstein equations in the ansatz for the metric used in \cite{Gr-Te}, the metric components end up having only $C^{0,1}$ regularity, {\it i.e.,} are Lipschitz continuous. In such a spacetime, the coordinate functions are $C^{1,1}$ giving the manifold this level of regularity. This paper investigates Riemannian geometry for $C^{1,1}$ spacetimes, the results of which may be useful in the study of weak solutions to the Einstein equations and numerical relativity. 

For $C^{1,1}$ spacetimes, an extra term arises in spacetime torsion leading to additional terms in the connection and curvature. Failing to account for these terms introduces a nonzero torsion into spacetime, affects covariant derivatives and curvature computations, and introduces an extraneous acceleration into particle paths. Correcting for theses terms requires a re-evaluation of Riemannian geometry from a foundational level. When appropriately accounted for, the assumption of $C^{1,1}$ regularity for spacetime leads to a geometry where the Ricci tensor is not symmetric.

\section{$C^{1,1}$ Riemannian Geometry}

Define the commutator between coordinate vector fields by 
\begin{equation}
[\partial_{\lambda},\partial_{\mu}]=\frac{\partial^{2}}{\partial x^{\lambda}\partial x^{\mu}}-\frac{\partial^{2}}{\partial x^{\mu}\partial x^{\lambda}}.\nonumber
\end{equation}
Normally, in differential geometry we require $[\partial_{\lambda},\partial_{\mu}]=0$ so that manifolds will have the $C^{2}$ differential structure of $\mathbb{R}^{n}$. Clairaut's Theorem says that mixed second-order partials commute provided the first-order partials exist and are continuous in an open set and the second-order partials are continuous at the point in question. However, this paper assumes that coordinate functions are only $C^{1,1}$ and hence a re-evaluation of the principles of Riemannian geometry is necessary. 

At this level of regularity, second-order partials may not commute. In order to apply Frobenius' Integrability Condition \cite{Lee}, we must assume that $[\partial_{\lambda},\partial_{\mu}]$ is in involution and hence 
\begin{equation}
[\partial_{\lambda},\partial_{\mu}]=c_{\lambda\mu}{}^{\sigma}\partial_{\sigma},
\label{Involution}
\end{equation}
where we have used the Einstein summation convention whereby repeated up and down indices are summed over all permissible values. Note that $c_{\lambda\mu}{}^{\sigma}=-c_{\mu\lambda}{}^{\sigma}$ and that these coefficients depend on the nature of the irregularities in question. The assumption that $c_{\lambda\mu}{}^{\sigma}\equiv 0$ is equivalent to assuming that spacetime coordinates are $C^{2}$. Also, we use a 3-index notation whereby objects that have two lower indices and one upper index and are skew in the lower indices have the upper index on the right, those symmetric in the lower indices have the upper index on the left, and those that are neither have their upper index in the middle.

Note that $c_{\lambda\mu}{}^{\sigma}$ is not a tensor since $[fX,Y]=f[X,Y]-Y(f)X$. 

\begin{definition}
The Jacobi tensor is defined by 
\begin{equation}
J(X,Y,Z)=[X,[Y,Z]]+[Z,[X,Y]]+[Y,[Z,X]].
\label{Jacobi}
\end{equation}
\end{definition}

That $J(X,Y,Z)$ is a tensor follows by a short computation, or as a consequence of equation (\ref{BiFirst}) from Theorem \ref{Bian}. Denote the components of this tensor by $J_{\alpha\beta\gamma}{}^{\sigma}$. 

\begin{theorem} The Jacobi tensor vanishes when $c_{\alpha\beta}{}^{\sigma}\equiv 0$ and 
\begin{equation}
J_{\alpha\beta\gamma}{}^{\sigma}=(c_{\alpha\beta}{}^{\sigma})_{,\gamma}+c_{\alpha\beta}{}^{\delta}c_{\gamma\delta}{}^{\sigma}+(c_{\gamma\alpha}{}^{\sigma})_{,\beta}+c_{\gamma\alpha}{}^{\delta}c_{\beta\delta}{}^{\sigma}+(c_{\beta\gamma}{}^{\sigma})_{,\alpha}+c_{\beta\gamma}{}^{\delta}c_{\alpha\delta}{}^{\sigma}.
\label{JIdent}
\end{equation}
\end{theorem}

{\bf Proof:} The proof follows by applying (\ref{Involution}) to the coordinate representation of (\ref{Jacobi}).\ $\Box{}$

The components of the connection are defined by 
\begin{equation}
\nabla_{\partial_{\lambda}}\partial_{\mu}=\Gamma_{\mu\ \lambda}^{\phantom{\mu}\sigma}\partial_{\sigma}
\label{ConnDef}
\end{equation}
where special note must be taken concerning index location and spacing.

\begin{theorem}So that covariant derivatives yield tensors, the components of the connection must transform according to
\begin{equation}
\Gamma_{i\ j}^{\ k}=\Gamma_{\mu\ \lambda}^{\phantom{\mu}\nu}\frac{\partial x^{\mu}}{\partial x^{i}}\frac{\partial x^{\lambda}}{\partial x^{j}}\frac{\partial x^{k}}{\partial x^{\nu}}-\frac{\partial x^{\mu}}{\partial x^{i}}\frac{\partial x^{\lambda}}{\partial x^{j}}\frac{\partial^{2} x^{k}}{\partial x^{\lambda}\partial x^{\mu}}.\label{ConnTrans}
\end{equation}
\end{theorem}

{\bf Proof:} This is a short computation.\ $\Box{}$

Note that since second-order partials do not commute, care must be taken with respect to the order of the indices in (\ref{ConnTrans}). Also, any object with two lower indices and one upper index that transforms according to the pattern in (\ref{ConnTrans}) is called a {\it connection}.

\begin{definition}
The torsion tensor is defined by 
\begin{equation}
T(X,Y)=\nabla_{X}Y-\nabla_{Y}X-[X,Y].
\label{TorDef}
\end{equation}
That (\ref{TorDef}) is a tensor follows by a short computation.
\end{definition}

The coordinates for the torsion tensor are given by 
\begin{equation}
T_{\lambda\mu}{}^{\nu}=\Gamma_{\mu\ \lambda}^{\ \nu}-\Gamma_{\lambda\ \mu}^{\ \nu}-c_{\lambda\mu}{}^{\nu}.
\label{TorComp}
\end{equation}

\begin{theorem} For $C^{1,1}$ torsion-free manifolds,
\begin{equation}
\Gamma_{\mu\ \lambda}^{\ \alpha}=\Gamma_{\lambda\ \mu}^{\ \alpha}+c_{\lambda\mu}{}^{\alpha}.
\label{ChSym}
\end{equation}
\end{theorem}

{\bf Proof:} Torsion-free connections satisfy
\begin{equation}
T(X,Y)=\nabla_{X}Y-\nabla_{Y}X-[X,Y]=0,
\nonumber
\end{equation}
which will be affected by the integrability conditions (\ref{Involution}). This means that 
\begin{eqnarray}
0&=&\nabla_{\partial_{\lambda}}\partial_{\mu}-\nabla_{\partial_{\mu}}\partial_{\lambda}-[\partial_{\lambda},\partial_{\mu}]\nonumber\\
&=&\Gamma_{\mu\ \lambda}^{\ \alpha}\partial_{\alpha}-\Gamma_{\lambda\ \mu}^{\ \alpha}\partial_{\alpha}-c_{\lambda\mu}{}^{\alpha}\partial_{\alpha}
\nonumber
\end{eqnarray}
and hence (\ref{ChSym}) holds.\ $\Box{}$

{\it Nota bene:} The components of the connection are not symmetric in their lower indices. Note also 
\begin{equation}
\Gamma_{\lambda\ \mu}^{\ \alpha}-\Gamma_{\mu\ \lambda}^{\ \alpha}=-c_{\lambda\mu}{}^{\alpha}.\label{ChDiff}
\end{equation}

A short computations gives the following theorem.
\begin{theorem}The quantities $c_{\lambda\mu}{}^{\nu}$ transform according to 
\begin{equation}
c_{ij}{}^{k}=c_{\alpha\beta}{}^{\nu}\frac{\partial x^{\alpha}}{\partial x^{i}}\frac{\partial x^{\beta}}{\partial x^{j}}\frac{\partial x^{k}}{\partial x^{\nu}}-\frac{\partial x^{\alpha}}{\partial x^{i}}\frac{\partial x^{\beta}}{\partial x^{j}}\left(\frac{\partial^{2} x^{k}}{\partial x^{\alpha}\partial x^{\beta}}-\frac{\partial^{2} x^{k}}{\partial x^{\beta}\partial x^{\alpha}}\right).\label{cTrans}
\end{equation}
\end{theorem}

\begin{definition} Set
\begin{eqnarray}
\left\{_{\lambda\ \mu}^{\ \nu}\right\}&=&\tfrac{1}{2}g^{\nu\alpha}\left(-g_{\lambda\mu,\alpha}+g_{\alpha \lambda,\mu}+g_{\mu\alpha,\lambda}\right)\label{UsualCr}\\
S_{\lambda\mu}{}^{\nu}&=&-\tfrac{1}{2}g^{\nu\alpha}c_{\lambda\mu}{}^{\sigma}g_{\alpha\sigma}=-\tfrac{1}{2}c_{\lambda\mu}{}^{\nu},\label{ConnSkew1}\\
U^{\nu}{}_{\lambda\mu}&=&\tfrac{1}{2}g^{\nu\alpha}\left(c_{\alpha \lambda}{}^{\sigma}g_{\sigma \mu}+c_{\alpha \mu}{}^{\sigma}g_{\lambda\sigma}\right).\label{ConnSym1}
\end{eqnarray}
The symbols $\left\{_{\lambda\ \mu}^{\ \nu}\right\}$ are called the Levi-Civita symbols. 
\end{definition}

Note that by (\ref{Involution}) the expression $S_{\lambda\mu}{}^{\nu}$ is not a tensor, and $U^{\nu}{}_{\lambda\mu}$ is not a tensor by a brief calculation.  

\begin{theorem} For $C^{1,1}$ Riemannian manifolds, the components of the connection are given by 
\begin{eqnarray}
\Gamma_{\lambda\ \mu}^{\ \nu}=\left\{_{\lambda\ \mu}^{\phantom{\lambda\,}\nu}\right\}+S_{\lambda\mu}{}^{\nu}+U^{\nu}{}_{\lambda\mu}.\label{ChrSym}
\end{eqnarray}
\end{theorem}
Note: Christoffel symbols are neither symmetric nor skew in their lower indices for $C^{1,1}$ Riemannian manifolds. 

{\bf Proof:} If the metric is parallel, then
\begin{eqnarray}
0&=&g_{\lambda\mu;\nu}\nonumber\\
&=&g_{\lambda\mu,\nu}-\Gamma_{\lambda\ \nu}^{\ \alpha}g_{\alpha\mu}-\Gamma_{\mu\ \nu}^{\ \alpha}g_{\lambda\alpha}.\nonumber
\end{eqnarray}
Via the usual Christoffel elimination, the components of the connection in terms of the metric and $c_{\lambda\mu}{}^{\sigma}$ are found. In particular,
\begin{eqnarray}
-g_{\lambda\mu,\nu}+g_{\nu\lambda,\mu}+g_{\mu \nu,\lambda}&=&-\Gamma_{\lambda\ \nu}^{\ \alpha}g_{\alpha \mu}-\Gamma_{\mu\ \nu}^{\ \alpha}g_{\lambda\alpha}+\Gamma_{\nu\ \mu}^{\ \alpha}g_{\alpha \lambda}+\Gamma_{\lambda\ \mu}^{\ \alpha}g_{\nu\alpha}\nonumber\\
&&+\Gamma_{\mu\ \lambda}^{\ \alpha}g_{\alpha \nu}+\Gamma_{\nu\ \lambda}^{\ \alpha}g_{\mu\alpha}.\label{Elim1}
\end{eqnarray}
Applying (\ref{ChDiff}) to (\ref{Elim1}) yields
\begin{equation}
-g_{\lambda\mu,\nu}+g_{\nu\lambda,\mu}+g_{\mu\nu,\lambda}=-c_{\nu\lambda}{}^{\alpha}g_{\alpha \mu}-c_{\nu\mu}{}^{\alpha}g_{\lambda\alpha}+c_{\lambda\mu}{}^{\alpha}g_{\nu\alpha}+2\Gamma_{\lambda\ \mu}^{\ \alpha}g_{\nu\alpha}\nonumber
\end{equation}
and hence (\ref{ChrSym}) holds. \ $\Box{}$

\begin{theorem}\label{BadConnThm} For $C^{1,1}$ manifolds the Levi-Civita symbols (\ref{UsualCr}) are not the components of a connection.
\end{theorem}

{\bf Proof:} The Levi-Civita symbols transform as follows:
\begin{eqnarray}
\left\{_{i\ j}^{\ k}\right\}&=&\tfrac{1}{2}g^{kl}\left(-g_{ij,l}+g_{li,j}+g_{jl,i}\right)\nonumber\\
&=&\tfrac{1}{2}\frac{\partial x^{k}}{\partial x^{\alpha}}\frac{\partial x^{l}}{\partial x^{\beta}}g^{\alpha\beta}\left(-\frac{\partial x^{\delta}}{\partial x^{l}}\frac{\partial }{\partial x^{\delta}}\left[g_{\lambda\mu}\frac{\partial x^{\lambda}}{\partial x^{i}}\frac{\partial x^{\mu}}{\partial x^{j}}\right]+\frac{\partial x^{\mu}}{\partial x^{j}}\frac{\partial }{\partial x^{\mu}}\left[g_{\delta\lambda}\frac{\partial x^{\delta}}{\partial x^{l}}\frac{\partial x^{\lambda}}{\partial x^{i}}\right]\right.\nonumber\\
&&\left.+\frac{\partial x^{\lambda}}{\partial x^{i}}\frac{\partial }{\partial x^{\lambda}}\left[g_{\mu\delta}\frac{\partial x^{\mu}}{\partial x^{j}}\frac{\partial x^{\delta}}{\partial x^{l}}\right]\right)\nonumber\\
&=&\left\{_{\lambda\ \mu}^{\phantom{\lambda}\nu}\right\}\frac{\partial x^{\lambda}}{\partial x^{i}}\frac{\partial x^{\mu}}{\partial x^{j}}\frac{\partial x^{k}}{\partial x^{\nu}}+\tfrac{1}{2}\frac{\partial x^{k}}{\partial x^{\alpha}}\frac{\partial x^{l}}{\partial x^{\beta}}g^{\alpha\beta}\left(-g_{\lambda\mu}\frac{\partial }{\partial x^{l}}\left[\frac{\partial x^{\lambda}}{\partial x^{i}}\frac{\partial x^{\mu}}{\partial x^{j}}\right]\right.\nonumber\\
&&\left.+g_{\delta\lambda}\frac{\partial }{\partial x^{j}}\left[\frac{\partial x^{\delta}}{\partial x^{l}}\frac{\partial x^{\lambda}}{\partial x^{i}}\right]+g_{\mu\delta}\frac{\partial }{\partial x^{i}}\left[\frac{\partial x^{\mu}}{\partial x^{j}}\frac{\partial x^{\delta}}{\partial x^{l}}\right]\right)\nonumber\\
&=&\left\{_{\lambda\ \mu}^{\phantom{\lambda}\nu}\right\}\frac{\partial x^{\lambda}}{\partial x^{i}}\frac{\partial x^{\mu}}{\partial x^{j}}\frac{\partial x^{k}}{\partial x^{\nu}}+\tfrac{1}{2}\frac{\partial x^{k}}{\partial x^{\alpha}}\frac{\partial x^{l}}{\partial x^{\beta}}g^{\alpha\beta}\left(-g_{\lambda\mu}\left[\frac{\partial^{2} x^{\lambda}}{\partial x^{l}\partial x^{i}}\frac{\partial x^{\mu}}{\partial x^{j}}+\frac{\partial x^{\lambda}}{\partial x^{i}}\frac{\partial^{2} x^{\mu}}{\partial x^{l}\partial x^{j}}\right]\right.\nonumber\\
&&+g_{\delta\lambda}\left[\frac{\partial^{2} x^{\delta}}{\partial x^{j}\partial x^{l}}\frac{\partial x^{\lambda}}{\partial x^{i}}+\frac{\partial x^{\delta}}{\partial x^{l}}\frac{\partial^{2} x^{\lambda}}{\partial x^{j}\partial x^{i}}\right]\nonumber\\
&&\left.+g_{\mu\delta}\left[\frac{\partial^{2} x^{\mu}}{\partial x^{i}\partial x^{j}}\frac{\partial x^{\delta}}{\partial x^{l}}+\frac{\partial x^{\mu}}{\partial x^{j}}\frac{\partial^{2} x^{\delta}}{\partial x^{i}\partial x^{l}}\right]\right)\nonumber\\
&=&\left\{_{\lambda\ \mu}^{\phantom{\lambda}\nu}\right\}\frac{\partial x^{\lambda}}{\partial x^{i}}\frac{\partial x^{\mu}}{\partial x^{j}}\frac{\partial x^{k}}{\partial x^{\nu}}+\tfrac{1}{2}\frac{\partial x^{k}}{\partial x^{\lambda}}\left[\frac{\partial^{2} x^{\lambda}}{\partial x^{j}\partial x^{i}}+\frac{\partial^{2} x^{\mu}}{\partial x^{i}\partial x^{j}}\right]+\nonumber\\
&&\tfrac{1}{2}\frac{\partial x^{k}}{\partial x^{\alpha}}\frac{\partial x^{l}}{\partial x^{\beta}}g^{\alpha\beta}\left(g_{\mu\delta}\frac{\partial x^{\mu}}{\partial x^{j}}\left(\frac{\partial^{2} x^{\delta}}{\partial x^{i}\partial x^{l}}-\frac{\partial^{2} x^{\delta}}{\partial x^{l}\partial x^{i}}\right)\right.\nonumber\\
&&\left.+g_{\delta\lambda}\frac{\partial x^{\lambda}}{\partial x^{i}}\left(\frac{\partial^{2} x^{\delta}}{\partial x^{j}\partial x^{l}}-\frac{\partial^{2} x^{\delta}}{\partial x^{l}\partial x^{j}}\right)\right).\label{BadTrans}
\end{eqnarray}
Due to the lack of commutativity in second-order partials, (\ref{BadTrans}) shows that the Levi-Civita symbols do not form a connection.\ $\Box{}$

\begin{theorem}For $C^{1,1}$ Riemannian manifolds, the symbols $\Gamma_{\lambda\ \mu}^{\ \nu}$ given by (\ref{ChrSym}) form the components of a connection.
\end{theorem}

{\bf Proof:} The proof follows from (\ref{BadTrans}), (\ref{cTrans}), (\ref{ConnSkew1}) and (\ref{ConnSym1}).\ $\Box{}$ 

If $T_{\lambda\mu\ldots\xi}$ is an object, let $E_{\lambda\mu\ldots\xi}$ be the sum of all $T$'s with even permutations of the indices and let $O_{\lambda\mu\ldots\xi}$ be the sum of all $T$'s with odd permutations. If there are $p$ indices, set
\begin{eqnarray}
T_{(\lambda\mu\ldots\xi)}&=&\tfrac{1}{p!}\left(E_{\lambda\mu\ldots\xi}+O_{\lambda\mu\ldots\xi}\right)\nonumber\\
T_{[\lambda\mu\ldots\xi]}&=&\tfrac{1}{p!}\left(E_{\lambda\mu\ldots\xi}-O_{\lambda\mu\ldots\xi}\right).\nonumber
\end{eqnarray}
To exclude indices, enclose them in $|\cdot|$.\footnote{For a detailed discussion of this notation, see \cite{Hlavaty}} For example,
\begin{equation}
T_{(\alpha|\beta|\gamma)}=\tfrac{1}{2}\left(T_{\alpha\beta\gamma}+T_{\gamma\beta\alpha}\right).
\nonumber
\end{equation}

Note that
\begin{eqnarray}
S_{\lambda\mu}{}^{\nu}&=&S_{[\lambda\mu]}{}^{\nu}=-\tfrac{1}{2}c_{\lambda\mu}{}^{\nu},\label{ConnSkew}\\
U^{\nu}{}_{\lambda\mu}&=&U^{\nu}{}_{(\lambda\mu)}=g^{\nu\alpha}c_{\alpha (\lambda}{}^{\sigma}g_{\mu)\sigma}=-2S^{\nu}{}_{(\lambda\mu)}.\label{ConnSym}
\end{eqnarray}
Also, (\ref{UsualCr}) and (\ref{ConnSym}) are symmetric in $\lambda$ and $\mu$ while (\ref{ConnSkew}) is skew. It follows that geodesic paths are affected by $U^{\nu}{}_{\lambda\mu}$ and through this term by $c_{\lambda\mu}{}^{\nu}$ but not by $S_{\lambda\mu}{}^{\nu}$ directly. To see this, note that geodesic paths satisfy 
\begin{eqnarray}
0&=&\frac{d^{2}x^{\nu}}{dq^{2}}+\Gamma_{\alpha\ \beta}^{\phantom{\alpha}\nu}\frac{d x^{\alpha}}{dq}\frac{d x^{\beta}}{dq}\nonumber\\
&=&\frac{d^{2}x^{\nu}}{dq^{2}}+\Gamma_{(\alpha\ \beta)}^{\phantom{(\alpha}\nu}\frac{d x^{\alpha}}{dq}\frac{d x^{\beta}}{dq}.\nonumber
\end{eqnarray}

We have 
\begin{equation}
\Gamma_{(\lambda\ \mu)}^{\phantom{(\lambda}\nu}=\left\{_{\lambda\ \mu}^{\ \nu}\right\}+U^{\nu}{}_{\lambda\mu}\nonumber
\end{equation}
and
\begin{equation}
\Gamma_{[\lambda\ \mu]}^{\phantom{[\lambda}\nu}=S_{\lambda\mu}{}^{\nu}.\nonumber
\end{equation}
Also, due to their symmetries,
\begin{eqnarray}
g_{\lambda\mu}&=&g_{(\lambda\mu)}\nonumber\\
c_{\lambda\mu}{}^{\nu}&=&c_{[\lambda\mu]}{}^{\nu}.\nonumber
\end{eqnarray}
We also define
\begin{eqnarray}
S_{\sigma}&=&S_{\sigma\alpha}{}^{\alpha}\nonumber\\
U_{\sigma}&=&U^{\alpha}{}_{\sigma\alpha}.\nonumber
\end{eqnarray}
Note that $S_{\sigma}=U_{\sigma}=-\tfrac{1}{2}c_{\sigma\alpha}{}^{\alpha}$.

\begin{theorem} We have
\begin{eqnarray}
\Gamma_{\alpha\ \sigma}^{\phantom{\sigma}\sigma}&=&\tfrac{1}{2}\ln(-g)_{,\alpha}-c_{\alpha\sigma}{}^{\sigma}=\tfrac{1}{2}\ln(-g)_{,\alpha}+2S_{\alpha}\nonumber\\
\Gamma_{\sigma\ \alpha}^{\phantom{\sigma}\sigma}&=&\tfrac{1}{2}\ln(-g)_{,\alpha}.\nonumber
\end{eqnarray}
\end{theorem}
The proof is obvious.

\begin{definition} The Riemann curvature tensor is defined by 
\begin{equation}
R(X,Y)Z=\nabla_{X}\nabla_{Y}Z-\nabla_{Y}\nabla_{X}Z-\nabla_{[X,Y]}Z.\label{RCurv}
\end{equation}
\end{definition}

\begin{theorem} In $C^{1,1}$ Riemannian manifolds, the Riemann curvature tensor is given by 
\begin{eqnarray}
R^{\xi}{}_{\nu\lambda\mu}&=&\Gamma_{\nu\ \mu,\lambda}^{\phantom{\nu\,}\xi}-\Gamma_{\nu\ \lambda,\mu}^{\phantom{\nu\,}\xi}+\Gamma_{\nu\ \mu}^{\phantom{\nu\,}\sigma}\Gamma_{\sigma\ \lambda}^{\phantom{\nu\,}\xi}-\Gamma_{\nu\ \lambda}^{\phantom{\nu\,}\sigma}\Gamma_{\sigma\ \mu}^{\phantom{\nu\,}\xi}-c_{\lambda\mu}{}^{\sigma}\Gamma_{\nu\ \sigma}^{\phantom{\nu\,}\xi}\nonumber\\
&=&2\left(\Gamma_{\nu\ [\mu,\lambda]}^{\phantom{\nu\,}\xi}+\Gamma_{\nu\ [\mu}^{\phantom{\nu\,}\sigma}\Gamma_{|\sigma|\ \lambda]}^{\phantom{\nu\,}\xi}\right)-c_{\lambda\mu}{}^{\sigma}\Gamma_{\nu\ \sigma}^{\phantom{\nu\,}\xi}.\label{RiemCur}
\end{eqnarray}
\end{theorem}

{\bf Proof:} The proof follows by substituting coordinates into (\ref{RCurv}).\ $\Box{}$

Note: There are differing conventions for the positions of the indices on the Riemann curvature tensor. The present convention was chosen for convenience. 

\begin{theorem} The Riemann curvature tensor has the symmetries
\begin{eqnarray}
R^{\xi}{}_{\nu\mu\lambda}&=&R^{\xi}{}_{\nu[\mu\lambda]}\label{Sym1}\\
R_{\xi\nu\mu\lambda}&=&R_{[\xi\nu]\mu\lambda}.\label{Sym2}
\end{eqnarray}
\end{theorem}

{\bf Proof:} Equation (\ref{Sym1}) is immediately apparent from (\ref{RiemCur}). Equation (\ref{Sym2}) follows from the definition (\ref{RCurv}) after a short computation.\ $\Box{}$

Equation (\ref{RiemCur}), noting (\ref{ChrSym}), involves partial derivatives of the functions $c_{\lambda\mu}{}^{\alpha}$ and hence must be interpreted in the sense of distributions.

\begin{theorem} We have
\begin{equation}
v^{\nu}{}_{;\mu;\omega}-v^{\nu}{}_{;\omega;\mu}=-R^{\nu}{}_{\sigma\mu\omega}v^{\sigma}.
\label{RCov}
\end{equation}
\end{theorem}

{\bf Proof:} Applying (\ref{RiemCur}) and (\ref{Involution}) yields (\ref{RCov}).\ $\Box{}$

\begin{definition} Ricci curvature, also called the first contracted curvature tensor, is given by 
\begin{eqnarray}
R_{\nu\mu}&\defeq&R^{\sigma}{}_{\nu\sigma\mu}\nonumber\\
&=&\Gamma_{\nu\ \mu,\sigma}^{\phantom{\nu}\sigma}-\Gamma_{\nu\ \sigma,\mu}^{\phantom{\nu}\sigma}+\Gamma_{\nu\ \mu}^{\phantom{\nu}\alpha}\Gamma_{\alpha\ \sigma}^{\phantom{\nu}\sigma}-\Gamma_{\nu\ \sigma}^{\phantom{\nu}\alpha}\Gamma_{\alpha\ \mu}^{\phantom{\nu}\sigma}-c_{\alpha\mu}{}^{\sigma}\Gamma_{\nu\ \sigma}^{\phantom{\nu}\alpha}.\nonumber
\end{eqnarray}
\end{definition}
Note that the Ricci tensor is not necessarily symmetric. 

\begin{theorem} The skew-symmetric components of the Ricci tensor vanish when $c_{\lambda\mu}{}^{\nu}\equiv 0$. In particular,
\begin{equation}
R_{[\nu\mu]}=(S_{\nu\mu}{}^{\sigma})_{,\sigma}-2S_{[\nu,\mu]}+2S_{\alpha}S_{\nu\mu}{}^{\alpha}.
\label{SkRia}
\end{equation}
\end{theorem}

{\bf Proof:} To see this, note that 
\begin{equation}
R_{[\nu\mu]}=(S_{\nu\mu}{}^{\sigma})_{,\sigma}-\Gamma_{[\nu\ |\sigma|,\mu]}^{\phantom{[\nu}\sigma}+S_{\nu\mu}{}^{\alpha}\Gamma_{\alpha\ \sigma}^{\phantom{\sigma}\sigma}-\Gamma_{[\nu\ |\alpha}^{\phantom{\sigma}\sigma}\Gamma_{\sigma|\ \mu]}^{\phantom{\sigma|}\alpha}-c_{\alpha[\mu}{}^{\sigma}\Gamma_{\nu]\ \sigma}^{\phantom{\nu]}\alpha},
\label{SkRi}
\end{equation}
so the skew-symmetric components of the Ricci tensor vanish provided we can show that the second and fourth terms on the right-hand-side of (\ref{SkRi}) vanish with $c_{\lambda\mu}{}^{\nu}$. Now,
\begin{eqnarray}
\Gamma_{[\nu\ |\sigma|,\mu]}^{\phantom{[\nu}\sigma}&=&\tfrac{1}{2}\left[\left(\tfrac{1}{2}\frac{g_{,\nu}}{g}+2S_{\nu}\right)_{,\mu}-\left(\tfrac{1}{2}\frac{g_{,\mu}}{g}+2S_{\mu}\right)_{,\nu}\right]\nonumber\\
&=&S_{\nu\mu}{}^{\alpha}\frac{g_{,\alpha}}{2g}+2S_{[\nu,\mu]}\nonumber
\end{eqnarray}
which vanishes with $c_{\lambda\mu}{}^{\nu}$. Finally, a short computation shows that 
\begin{equation}
-\Gamma_{[\nu\ |\alpha}^{\phantom{\sigma}\sigma}\Gamma_{\sigma|\ \mu]}^{\phantom{\sigma|}\alpha}=c_{\sigma[\mu}{}^{\alpha}\Gamma_{\nu]\ \alpha}^{\phantom{\mu}\sigma}\nonumber
\end{equation}
from which it follows that the skew components of the Ricci tensor vanish with $c_{\lambda\mu}{}^{\nu}$.

Synthesizing these computations yields (\ref{SkRia}).\ $\Box{}$

\begin{definition} The second contracted curvature tensor is given by 
\begin{equation}
V_{\mu\lambda}\defeq R^{\sigma}{}_{\sigma\mu\lambda}.\nonumber
\end{equation}
\end{definition}

\begin{theorem} $V_{\mu\lambda}=0$
\end{theorem}

{\bf Proof:} The proof follows from (\ref{Sym2}).\ $\Box{}$

\begin{theorem}\label{Bian}[The First Bianchi Identity] We have
\begin{equation}
R(X,Y)Z+R(Z,X)Y+R(Y,Z)X=J(X,Y,Z)\label{BiFirst}
\end{equation}
or
\begin{equation}
3R^{\xi}{}_{[\nu\lambda\mu]}=R^{\xi}{}_{\nu\lambda\mu}+R^{\xi}{}_{\mu\nu\lambda}+R^{\xi}{}_{\lambda\mu\nu}=J_{\nu\mu\lambda}{}^{\xi}\label{Bi1}
\end{equation}
where the right-hand-side is zero when $c_{\lambda\mu}{}^{\alpha}\equiv 0$.
\end{theorem}

{\bf Proof:} The proof makes essential use of the torsion-free property of Riemannian manifolds. We may write 
\begin{eqnarray}
R(X,Y)Z&+&R(Z,X)Y+R(Y,Z)X=\nabla_{X}\nabla_{Y}Z-\nabla_{Y}\nabla_{X}Z-\nabla_{[X,Y]}Z\nonumber\\
&&+\nabla_{Z}\nabla_{X}Y-\nabla_{X}\nabla_{Z}Y-\nabla_{[Z,X]}Y\nonumber\\
&&+\nabla_{Y}\nabla_{Z}X-\nabla_{Z}\nabla_{Y}X-\nabla_{[Y,Z]}X\nonumber\\
&=&\nabla_{X}[Y,Z]-\nabla_{[X,Y]}Z+\nabla_{Z}[X,Y]-\nabla_{[Z,X]}Y\nonumber\\
&&+\nabla_{Y}[Z,X]-\nabla_{[Y,Z]}X\nonumber\\
&=&[X,[Y,Z]]+[Z,[X,Y]]+[Y,[Z,X]]\nonumber\\
&=&J(X,Y,Z). \ \Box{}\nonumber
\end{eqnarray}

\begin{theorem}[The Second Bianchi Identity] We have
\begin{equation}
\nabla_{W}(R(X,Y)Z)+\nabla_{Y}(R(W,X)Z)+\nabla_{X}(R(Y,W)Z)=-\nabla_{J(X,Y,W)}Z\label{B2}
\end{equation}
or
\begin{equation}
3R^{\xi}{}_{\nu[\lambda\mu;\eta]}=R^{\xi}{}_{\nu\lambda\mu;\eta}+R^{\xi}{}_{\nu\eta\lambda;\mu}+R^{\xi}{}_{\nu\mu\eta;\lambda}=-J_{\mu\lambda\eta}{}^{\sigma}\Gamma_{\nu\ \sigma}^{\phantom{\eta}\xi}
\label{Bi2}
\end{equation}
where the right-hand-side is zero when $c_{\lambda\mu}{}^{\alpha}\equiv 0$.
\end{theorem}

{\bf Proof:} Application of (\ref{Involution}), (\ref{RiemCur}) and (\ref{Jacobi}) leads to (\ref{Bi2}). \ $\Box{}$

\begin{definition} Set
\begin{equation}R=R_{\nu\mu}g^{\nu\mu}.
\nonumber
\end{equation}
\end{definition}
Note that $R=R_{(\nu\mu)}g^{\nu\mu}$.

\begin{definition} The Einstein tensor is defined by 
\begin{equation}
G_{\mu\nu}=R_{\mu\nu}-\tfrac{1}{2}R g_{\mu\nu}.
\nonumber
\end{equation}
\end{definition}

\begin{theorem} The divergence of the Einstein tensor vanishes when $c_{\nu\mu}{}^{\lambda}\equiv 0$ and
\begin{equation}
{\rm div}G_{\mu\cdot}=g^{\nu\lambda}G_{\mu\nu;\lambda}=-\tfrac{1}{2}g^{\nu\lambda}J_{\mu\lambda\alpha}{}^{\sigma}\Gamma_{\nu\ \sigma}^{\phantom{\eta}\alpha}.
\label{ConsEq}
\end{equation}
\end{theorem}

{\bf Proof:} Applying (\ref{Bi2}) and the curvature symmetries, we have
\begin{eqnarray}
-J_{\mu\lambda\alpha}{}^{\sigma}\Gamma_{\nu\ \sigma}^{\phantom{\eta}\alpha}&=&R^{\sigma}{}_{\nu\mu\lambda;\sigma}+R_{\nu\lambda;\mu}-R_{\nu\mu;\lambda}\nonumber\\
&=&-g^{\sigma\alpha}g_{\nu\beta}R^{\beta}{}_{\alpha\mu\lambda;\sigma}+R_{\nu\lambda;\mu}-R_{\nu\mu;\lambda}.\nonumber
\end{eqnarray}
Contracting on $\nu$ and $\lambda$ yields
\begin{eqnarray}
-g^{\nu\lambda}J_{\mu\lambda\alpha}{}^{\sigma}\Gamma_{\nu\ \sigma}^{\phantom{\eta}\alpha}&=&-g^{\nu\lambda}g^{\sigma\alpha}g_{\nu\beta}R^{\beta}{}_{\alpha\mu\lambda;\sigma}+R_{,\mu}-g^{\nu\lambda}R_{\nu\mu;\lambda}\nonumber\\
&=&-g^{\sigma\alpha}R_{\alpha\mu;\sigma}+R_{,\mu}-g^{\nu\lambda}R_{\nu\mu;\lambda}\nonumber\\
&=&R_{,\mu}-2g^{\nu\lambda}R_{\nu\mu;\lambda}\nonumber
\end{eqnarray}
from which (\ref{ConsEq}) follows.\ $\Box{}$

From (\ref{ConsEq}) we see that the Einstein tensor is not divergence-free. This would remain true even if we had defined $G_{\mu\nu}=R_{(\mu\nu)}-\tfrac{1}{2}R g_{\mu\nu}$ and in this case its divergence would still vanish with $c_{\nu\mu}{}^{\lambda}$ in light of (\ref{SkRia}). 

\section{Conclusion} Differential geometry of spacetimes having $C^{1,1}$ coordinates differs in important ways from the usual differential geometry. The commutator of coordinate vector fields may not vanish but must be assumed to be in involution. This introduces new terms into the connection through the assumption that spacetime is torsion-free. Failing to account for these terms introduces nonzero torsion into spacetime. The new symmetric terms in the modified connection affect geodesic paths. In addition, the connection on $C^{1,1}$ spacetimes introduces new terms into the curvature leading to alterations of the Bianchi identities. The Ricci tensor ceases to be symmetric and the Einstein tensor ceases to be divergence-free. It is essential that these methods be taken into account for manifolds having only $C^{1,1}$ regularity.

\end{document}